\documentclass[pre,twocolumn,showpacs]{revtex4}

\usepackage{graphicx}
\usepackage{amsmath}
\usepackage{mathtools}

\newcommand{\be}{\begin{equation}}
\newcommand{\ee}{\end{equation}}
\newcommand{\ba}{\begin{eqnarray}}
\newcommand{\ea}{\end{eqnarray}}

\begin{document}

\title{The double-layer structure of overscreened surfaces by smeared-out ions
}

\author{Derek Frydel}
\email{dfrydel@gmail.com}
\affiliation{Institute for Advanced Study, Shenzhen University, Shenzhen, Guangdong 518060, China}

\date{\today}

\begin{abstract}
The present work focuses on the structure of a double-layer of overscreened charged surfaces by smeared-out 
charges and probes the link between the structure of a double-layer and the bulk properties of an electrolyte with 
special view to the role of the Kirkwood crossover.  Just as the Kirkwood line divides a bulk solution into a fluid
with monotonic and oscillatory decaying correlations, it similarly separates charge inversion into two broad domains, 
with and without oscillating charge density profile.  As initially oscillations may appear 
like a far-field occurrence, eventually they develop into a full fledged layering of a charge density.  
\end{abstract}

\pacs{
}

\maketitle

\section{Introduction}

In our previous work we reported the possibility of charge inversion for ions with charge distribution 
$w(r)\ne\delta(r)$ \cite{Frydel13,Frydel16} (the so called smeared-out ions).  Because the phenomenon 
was captured within the mean-field model, correlations have no role in this behavior.  
Rather the explanation lies in 
reduced interactions at short separations due to elimination of a divergence in the Coulomb 
functional form, which permit more compact structure of counterions.  This is in contrast to point charges, 
$w(r)=\delta(r)$, where charge inversion 
occurs when strong correlations give rise to lateral ordering of counterions within a double-layer.  Such a correlation-based 
mechanism can only arise within an improved theory beyond the mean-field 
 \cite{Tellez05,Tellez06,Samaj06,Frydel16b}.

In Refs. \cite{Frydel13,Frydel16} we laid primary stress on the fact that a charge density profile (and an electrostatic potential) 
changes sign as a function of distance from a charged surface 
and gave less attention to the far-field region, especially the presence in that region of oscillations.  Our 
interest in the far-field region was recently revived by a demonstration in Ref. \cite{Hansen12} of the existence of the Kirkwood 
crossover -- a point where a charge-charge correlation function changes from a monotonically to an oscillatorily decaying 
profile -- and a possible connection to our own results in Ref. \cite{Frydel13}.  As oscillations in charge 
density profile around zero imply charge inversion 
(or rather an entire sequence of inversions), we began to wonder about a 
possible role of the Kirkwood crossover in the mechanism of charge inversion.  
If indeed 
charge inversion is coextensive with the onset of oscillatory decay, one may conclude that 
inversion is triggered by the Kirkwood crossover, and, therefore, is completely determined by the properties of a bulk 
electrolyte and independent of a charged surface.  If, however, there exists a possibility of charge inversion 
for monotonically decaying profile, then the picture is more subtle. 

The aim of the present work is to investigate charge inversion of smeared-out ions with special attention to 
the far-field region and the connection with the Kirkwood crossover.  Before 
proceeding, however, we review some basic ideas that will 
sketch a larger context and help to understand our motivations and perplexities about the far-field region.  

A characteristic of a fluid is a short-range translational order whereby correlations of a fluctuating 
variable decay exponentially as $f(r)e^{-r/\xi}$, where $\xi$ implies the correlation length and $f(r)$ 
is some algebraic function.  Now, if we turn to an inhomogeneous fluid, we find that a density perturbation 
falls off exponentially as it merges with a uniform fluid, $g(r)e^{-r/\xi}$, with the same exponential decay as that 
for correlations.  ($g(r)$ depends on the geometry of a
perturbation).  This universality of an exponential decay is a consequence of 
the linear response theory wherein a perturbation is expressed as a superposition of 
fluctuations.  Nonlinear effects of a larger perturbation do not invalidate this interpretation 
as in the region where a perturbation merges with a bulk fluid a perturbation is again small 
and the linear regime recovered.

A picture becomes more interesting if one considers dense fluids where the short-range translational 
order attains some degree of organization and the profile of a density-density correlation function develops 
a regular oscillating structure, albeit, that still decays exponentially.  There are now two parameters 
describing a decay of correlations: a correlation lenght $\xi$ and a wavelength $\lambda$.   The point 
where oscillations first appear is known as the Fisher-Widom crossover \cite{Evans94a,Evans01}.  It is a 
crossover in a sense that it does not entail any macroscopic changes nor generate discontinuities in 
thermodynamic quantities.  Discontinuous behavior, however, is observed in the scaling of the 
correlation length $\xi$:  
before the Fisher-Widom line $\xi$ decreases as a function of increasing density, 
after the crossover the trend changes and $\xi$ increases with a denser fluid.  
Now, if we turn to inhomogeneous fluids, we find a similar structure in a density decay:  an exponentially 
decaying oscillatory profile. 

An analogous picture emerges for electrolytes, except now the relevant quantities are a charge density and a 
charge-charge correlation function.  Here we also encounter a crossover from a monotonic to oscillatory decay 
that goes by the name of the Kirkwood crossover \cite{Kirkwood36,Evans94,Kjellander98}.  An interesting 
consequence of having oscillations in a charge density around zero is a repetitive charge inversion.  This 
type of charge inversion depends only on the conditions of a bulk that control the quantities 
$\xi$ and $\lambda$ and cannot be induced by a surface charge or any specific feature of a perturbation.  

On the other hand, we have a more "conventional" mechanism for charge inversion, 
wherein charge inversion is regarded as an effect
of a renormalized effective charge and does not assume any change in the far-field decay, which remains
monotonic.  (The idea of an "effective charge" entails the idea of a "dressed surface", a surface comprised of 
a bare surface charge plus an adjacent layer of counterions that together reduce an absolute value of a bare 
charge).  As the present work is entirely devoted to the mean-field analysis, any 
renormalization of an effective charge comes from nonlinear contributions of 
the mean-field theory, wherein the scaling of an effective charge corresponds to the scaling of a magnitude 
of a monotonically decaying profile.   
Such nonlinear renormalization of an effective charge carried out for point-ions does not lead to charge inversion but merely to 
charge saturation.  To capture charge inversion for point-ions one needs to go beyond the mean-field level 
of description \cite{Tellez05,Tellez06,Samaj06,Frydel16}.



This work is organized as follows.  
In section \ref{sec:sec2} we formulate the mean-field theory for smeared-out ions.  In section \ref{sec:sec3} we carry 
out the linear analysis to obtain the screening parameters and the location of the Kirkwood crossover for Gaussian 
distributed charges.  In section \ref{sec:sec4} we consider the full mean-field theory and focus on the renormalization 
of an effective surface charge for monotonically decaying profiles (prior to the Kirkwood crossover) and mark the region 
where charge inversion due to renormalization becomes possible.  In section \ref{sec:dumbbell} we make connection 
to a related system of dumbbell ions.  
Finally, in section \ref{sec:conclusion} we conclude 
the work.

\section{The mean-field description}
\label{sec:sec2}
Within the standard representation, an electrolyte is made up of $K$ different species of 
point-charges, and the Poisson equation is given by
\be
\epsilon\nabla^2\psi({\bf r}) = -\sum_{i=1}^Kq_i\rho_i({\bf r})
\ee
where $\psi({\bf r})$ is the electrostatic potential, $\rho_i({\bf r})$ and $q_i$ is a number density and 
a charge of a species $i$, respectively, and $\epsilon$ is the dielectric constant of a solvent medium.  
If we move away from a point-charge description and consider ions represented as smeared-out charges 
within a spherically symmetrical distribution $w_i({\bf r}-{\bf r}')$ and normalized as 
$\int d{\bf r}'\,w_i({\bf r}-{\bf r}')=1$, then the Poisson equation is 
\be
\epsilon\nabla^2\psi({\bf r}) = -\sum_{i=1}^Kq_i\int d{\bf r}'\,\rho_i({\bf r}')w_i({\bf r}-{\bf r}'),
\label{eq:poisson2}
\ee
and the case $w_i({\bf r}-{\bf r}')=\delta({\bf r}-{\bf r}')$ recovers the Poisson equation for point charges.
Note that since an ion has an extension, it does not need to be at ${\bf r}$ to contribute to 
a charge density at ${\bf r}$.  Contributions
come from a distribution $w(r)$ centered at ${\bf r}$, leading to a nonlocal term.  

The mean-field expression for a number density of smeared-out ions is 
\be
\rho_i({\bf r}) = c_ie^{-\beta q_i\int d{\bf r}'\,\psi({\bf r}')w_i({\bf r}-{\bf r}')},   
\ee
where $c_i$ in is the bulk concentration of a species $i$, and $\beta=1/k_BT$.  The nonlocality reflects the idea of an entire
ion $w(r)$ interacting  
with an electrostatic potential $\psi({\bf r})$.

Inserting the mean-field number density into the Poisson equation of Eq. (\ref{eq:poisson2}) yields a modified 
Poisson-Boltzmann equation for smeared-out ions, 
\be
\epsilon\nabla^2\psi({\bf r}) = \!-\!\!\sum_{i=1}^Kc_iq_i\!\!\int \!\!d{\bf r}' w_i({\bf r}-{\bf r}') 
e^{-\beta q_i\!\!\int d{\bf r}''\psi({\bf r}'')w_i({\bf r}'-{\bf r}'')}.  
\label{eq:MPB}
\ee

Not surprisingly, correlational contributions diminish with increasing size of an ion.  Consequently, 
the mean-field theory becomes virtually an exact model for representing ions with broad charge 
distributions.  
Our interest lies specifically in such a mean-field regime, where correlational contributions are minimal, so 
that any deviations from the point-charge model can be exclusively attributed to finite 
size.

\section{Linear analysis} 
\label{sec:sec3}
As mentioned in the introduction, to determine an asymptotic region within any theory (here we have in mind 
screening parameters or a correlation length) the linear analysis is sufficient.  Nonlinear 
contributions of the full mean-field model do not modify these parameters, while the linearized version of any theory
is simpler and easier to handle.  

Keeping only the linear terms of the modified PB equation in Eq. (\ref{eq:MPB}) we get 
\be 
\epsilon\nabla^2\psi({\bf r}) = \beta\sum_{i=1}^Kc_iq_i
\int \!\!d{\bf r}'\psi({\bf r}')\!\int \!\!d{\bf r}''w_i({\bf r}-{\bf r}'')w_i({\bf r}'-{\bf r}'').
\label{eq:lpb_sphere}
\ee
For a symmetric monovalent electrolyte, where the net charge of an ion is $q=\pm e$ ($e$ is the fundamental charge), 
the distributions are $w_{\pm}({\bf r}-{\bf r}')=\pm w({\bf r}-{\bf r}')$, and $c_s$ is the salt concentration in a bulk, 
the above equation further simplifies as 
\be
\epsilon\nabla^2\psi({\bf r}) = 2\beta c_s e^2
\int \!\!d{\bf r}'\psi({\bf r}')\!\int \!\!d{\bf r}''w({\bf r}-{\bf r}'') w({\bf r}''-{\bf r}').  
\ee

We perturb the uniform system by fixing a single ion at the origin, 
then obtain an electrostatic potential from the linear theory, 
\ba
\epsilon\nabla^2\psi(r) 
&=& 2\beta e^2c_s\!\!\int \!\!d{\bf r}'\psi(r')\!\!\!\int \!\!d{\bf r}''w({\bf r}-{\bf r}'') w({\bf r}''-{\bf r}')\nonumber\\
&-&ew(r).
\label{eq:lpb_sphere_11}
\ea
Note that a fixed particle is taken to be positive.  
Fourier transforming the above equation yields
\be
-\epsilon k^2\psi(k) = 2\beta e^2 c_s\psi(k)w^2(k) - ew(k),
\label{eq:lpb}
\ee
and the potential in the Fourier space is
\be
\psi(k) = \frac{ew(k)}{\epsilon k^2+2\beta e^2 c_sw^2(k)}.
\ee
At this point we introduce more convenient reduced units, 
\be
\phi(k) = \frac{4\pi\lambda_B w(k)}{k^2+\kappa_D^2w^2(k)}
\ee
where $\phi=\beta e\psi$, $\lambda_B=\beta e^2/(4\pi\epsilon)$ is the Bjerrum length, and 
$\kappa_D^2=8\pi c_s\lambda_B$ is the Debye screening parameter.  
The electrostatic potential in real space is then obtained from inverse Fourier transform, 
\be
\phi(r) = \frac{2\lambda_B}{\pi}\int_{0}^{\infty}dk\,\frac{k^2 w(k)}{k^2+\kappa_D^2w^2(k)}\frac{\sin kr}{kr}.  
\label{eq:phi}
\ee
For the case of point charges $w(k)=1$ and the above integral evaluates to a familiar screened potential
$\phi(r)=\lambda_B e^{-\kappa_D r}r^{-1}$.  

For particles with an arbitrary distribution $w(r)$, the integral in Eq. (\ref{eq:phi}) can be conveniently handled
using the residue theorem.  To adopt the method to the present problem, we alter the limits of the integration as
\be
\phi(r) = \frac{\lambda_B}{\pi i}\int_{-\infty}^{\infty}dk\,\frac{k w(k)}{k^2+\kappa_D^2w^2(k)}\frac{e^{ikr}}{r}.  
\label{eq:phi_2}
\ee
which is allowed as long as $w(k)$ is an even function, in which case the imaginary part cancels out.  The need 
to alter the integration limits will become clear as we outline the details of the method.  

In complex analysis the value of an integral along a closed curve $C$ can be expressed as 
a sum of residues inside the region enclosed by $C$, 
\be
\frac{1}{2\pi i}\oint_C dk\,f(k)  = \sum_{n}\text{Res}(f,k_n).
\label{eq:Res}
\ee
In this case $k$ is a complex variable, and $\text{Res}(f,k_n)$ is a residue of $f(k)$ at a pole $k_n$.  
A value of a residue corresponds to a coefficient $a_{-1}$ in the expansion
\be
f(k) = \sum_{m=-m_0}^{\infty}a_m(k-k_n)^m
\ee 
carried out in the neighborhood of a pole $k_n$.  A pole is said to be simple if $m_0=1$.  

In order to use Eq. (\ref{eq:Res}) to evaluated the integral in Eq. (\ref{eq:phi_2}), the curve $C$ should 
incorporate the real axis, while the integral along the remaining curve (let's say a half circle with radius $R\to\infty$) should 
evaluate to zero.  If satisfied, then a potential can be represented as
\be
\phi(r) = \frac{\lambda_B}{2\pi i}\oint_C dk\,f(k)  = \lambda_B\sum_{n}\text{Res}(f,k_n),
\label{eq:phi_3}
\ee
with the integrand $f(k)$ given by
\be
f(k) = \frac{2k w(k)}{k^2+\kappa_D^2w^2(k)}\frac{e^{ikr}}{r}.  
\ee

Poles, being singularities of the complex plane, correspond to zeros of the denominator
of $f(k)$, 
\be
k_n^2+\kappa_D^2w^2(k_n) = 0.  
\ee
If poles enclosed by $C$ are simple, and by representing $f(k)$ as a quotient of two functions 
${g(k)}/{h(k)}$, 
the residues are given by 
\be
\text{Res}(f,k_n) = \frac{g(k_n)}{h'(k_n)}. 
\ee
Together with Eq. (\ref{eq:phi_2}) and Eq. (\ref{eq:Res}), an electrostatic potential is then given by 
\be
\phi(r) = \lambda_B\sum_{n}\frac{e^{ik_nr}}{r}\frac{k_nw(k_n)}{k_n+\kappa_D^2w(k_n)w'(k_n)}.  
\label{eq:phi_3}
\ee

It is now clear that the poles, expressed as, 
\be
k_n = i\kappa_n + \omega_n,
\ee
characterize a screening parameter $\kappa_n$ and a wavenumber $\omega_n$ of each term in 
Eq. (\ref{eq:phi_3}).  Since the linear theory accurately describes only a far-field region, we are only 
interested in the dominant term, that is, the pole with the smallest $\kappa_n$.
 
Note that only a strictly imaginary $k_n$ yields a monotonically decaying function.  On the other hand, a 
strictly real $k_n$ yields a solid like structure with a long-range translational order.  A fully complex pole 
determines an oscillating exponentially decaying profile.

\subsection{$w(r)$ as a Gaussian distributed function}
As a specific case, we consider ions with a Gaussian distributed charge, 
\be
w(r) = \frac{e^{-r^2/2\sigma^2}}{(2\pi\sigma^2)^{3/2}},
\ee
whose Fourier transform is
\be
w(k) = e^{-k^2\sigma^2/2}, 
\ee
and an electrostatic potential within a linearized mean-field theory is given by
\be
\phi(r) = \frac{\lambda_B}{\pi i}\int_{-\infty}^{\infty}dk\,\frac{k e^{-k^2\sigma^2/2}}{k^2+\kappa_D^2e^{-k^2\sigma^2}}\frac{e^{ikr}}{r}. 
\ee
The poles satisfy 
\be
k_n^2+\kappa_D^2 e^{-k_n^2\sigma^2} = 0, 
\ee
or, after rearrangement, 
\be
-\kappa_D^2\sigma^2 = k_n^2\sigma^2 e^{k_n^2\sigma^2},
\ee
where $-\kappa_D^2\sigma^2$ appears as a function of $k_n^2\sigma^2$.  But being interested in
$k_n^2\sigma^2$ as a function of $-\kappa_D^2\sigma^2$, we look for an inverted relation that, in fact, 
is provided by the Lambert multivalued function \cite{Corless96,Dence13}
\be
W_n(-\kappa_D^2\sigma^2) = k_n^2\sigma^2,
\ee
where $n$ denotes a particular branch, with $n=0$ being the principal branch.  The Lambert function is the inverse 
relation of the function $f(W)=We^W$.  It turns out that this function is quite ubiquitous in nature.  For example, it provides 
an exact solution to the quantum-mechanical double-well Dirac delta function model for equal charges \cite{Corless96}.  
 
The poles are now expressed as
\be
k_n\sigma = i\sqrt{-W_n\big(-\kappa_D^2\sigma^2\big)}, 
\ee
and the electrostatic potential within the linear mean-field theory is
\be
\phi(r) = \sum_{n=-\infty}^{\infty}\frac{\lambda_Be^{ik_n r}}{r} \frac{e^{-k_n^2\sigma^2/2}}{1-k_n^2\sigma^2}.
\label{eq:phi_sum}
\ee

\begin{figure}[h] 
 \begin{center}%
 \begin{tabular}{rr}
  \includegraphics[height=0.25\textwidth,width=0.3\textwidth]{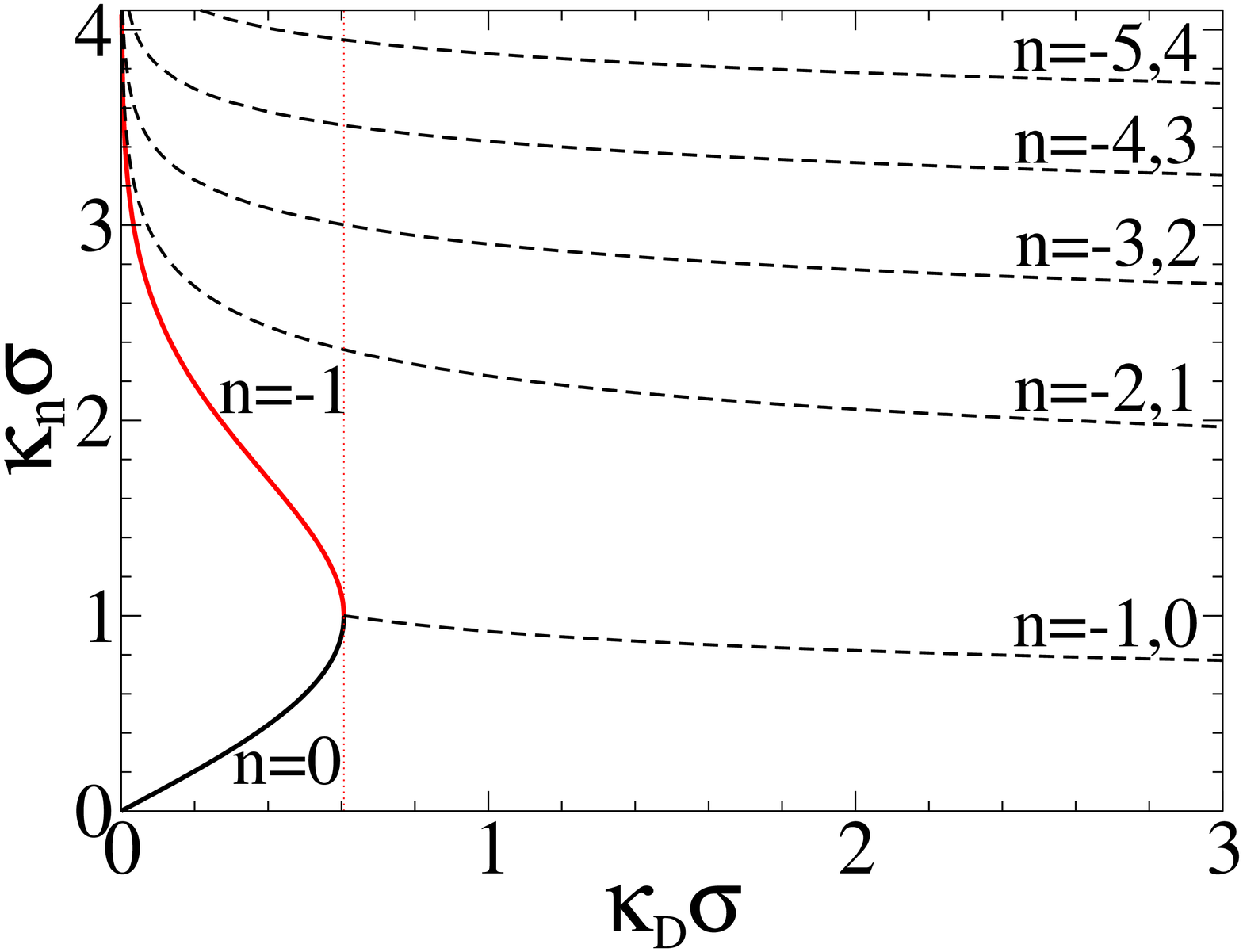}\\
\end{tabular}
 \end{center}
\caption{Screening parameters $\kappa_n$ for a number of initial branches for Gaussian distributed ion.  
The solid lines indicate the screening of monotonic and the dashed lines of oscillatory terms.  For 
$\kappa_D\sigma>e^{-1/2}$ all terms have oscillations and the point $\kappa_D\sigma=e^{-1/2}$ 
(a vertical dotted line) is represents the Kirkwood crossover.  }
\label{fig:kappa_n}
\end{figure}

In Fig. (\ref{fig:kappa_n}) we plot the screening parameters $\kappa_n$, obtained from poles, $k_n=i\kappa_n+\omega_n$, 
and given by
\be
\kappa_n = \text{Re}\bigg[\sqrt{-W_n\big(-\kappa_D^2\sigma^2\big)}\bigg], 
\ee
for a number of initial branches as a function of $\kappa_D\sigma$.  The Kirkwood crossover occurs at 
$\kappa_D\sigma=e^{-1/2}$.  At the crossover there is a discontinuity in the
scaling behavior for $\kappa_0$ and $\kappa_{-1}$, the two lowest branches that determine decay.  
\graphicspath{{figures/}}

\begin{figure}[h] 
 \begin{center}%
 \begin{tabular}{rr}
  \includegraphics[height=0.25\textwidth,width=0.3\textwidth]{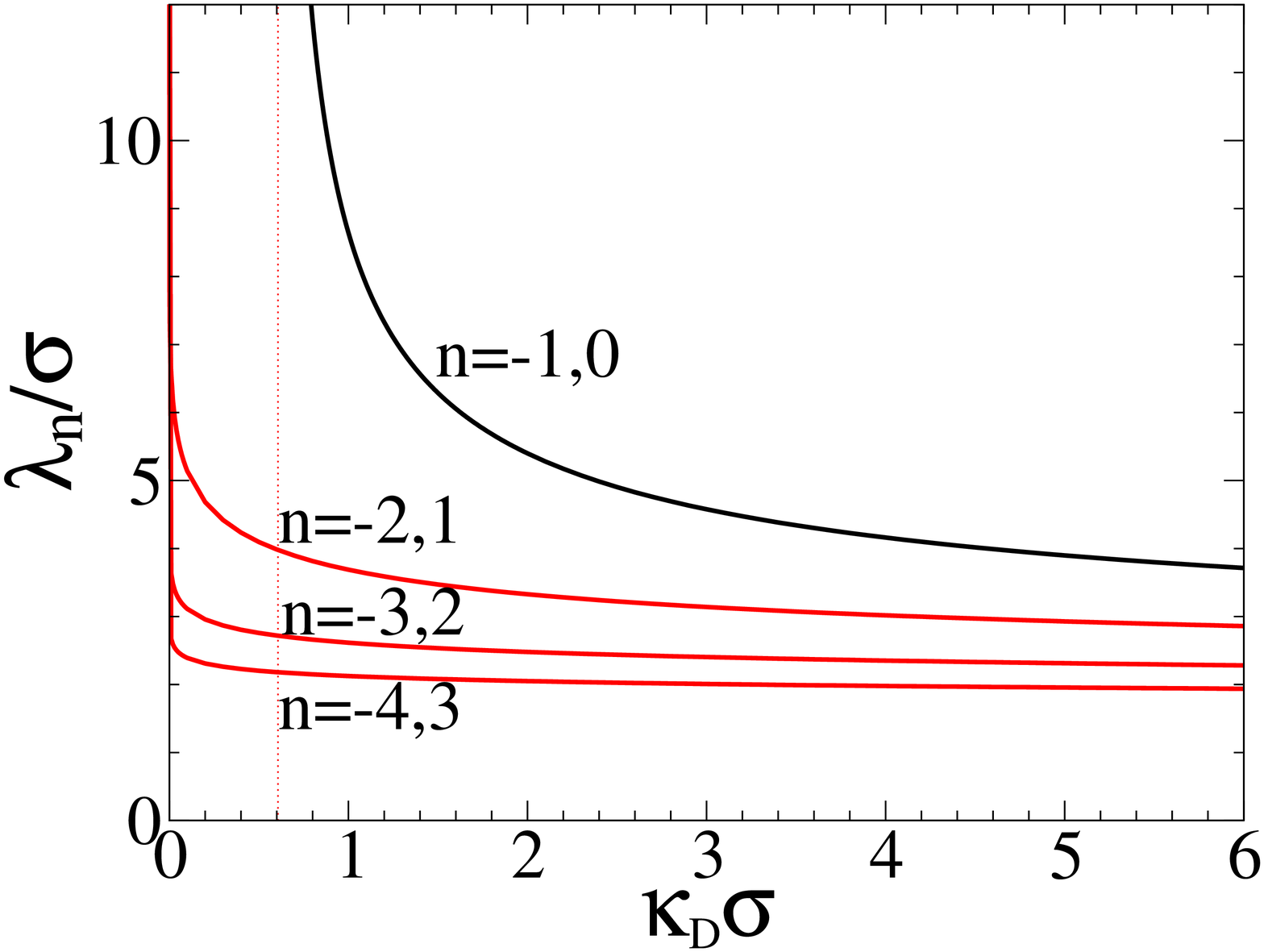}\\
\end{tabular}
 \end{center}
\caption{Wavelengths $\lambda_n=\frac{2\pi}{\omega_n}$ for a number of initial branches as a function of 
$\kappa_D\sigma$.  
$\lambda_0$ (which is equivalent with $\lambda_{-1}$) diverges at the Kirkwood crossover and the remaining 
wavelengths diverge as $\kappa_D\sigma\to 0$.  }
\label{fig:lambda_n}
\end{figure}

In Fig. (\ref{fig:lambda_n}) we plot the wavelengths $\lambda_n=2\pi/\omega_n$ where 
\be
\omega_n = \text{Im}\bigg[\sqrt{-W_n\big(-\kappa_D^2\sigma^2\big)}\bigg], 
\ee
for a number of initial branches.  
At the Kirkwood crossover, as $\kappa_D\sigma$ approaches $e^{-1/2}$ from above, $\lambda_0$ (and $\lambda_{-1}$) diverges, 
indicating the absence of oscillations for $\kappa_D\sigma<e^{-1/2}$.  
All the remaining $\lambda_n$ diverge only as $\kappa_D\sigma\to 0$.  
\graphicspath{{figures/}}

The crucial result of this section is the location of the Kirkwood crossover at $\kappa_D\sigma=e^{-1/2}$ 
and the precise determination of parameters governing the far-field decay, $\kappa_0$ and $\lambda_0$.  
The Kirkwood crossover splits an electrolyte into two domains.  Beyond the crossover the far-field decay of a 
charge density profile changes from monotonic to oscillatory.  The onset of oscillations in a charge density 
necessarily implies a repetitive charge inversion.  Such charge inversion is fundamentally different from a more 
conventional charge inversion that results from renormalization of an effective charge, firstly, because of its 
oscillating nature, and, secondly, because it depends on bulk properties alone.   
In contrast, conventional charge inversion happens for monotonically decaying profiles (prior to the Kirkwood 
crossover) and is triggered by strong correlations between counterions near a charged surface, therefore, 
the magnitude of a surface charge plays an important role \cite{Tellez05,Tellez06,Samaj06}.

\begin{figure}[h] 
 \begin{center}%
 \begin{tabular}{rr}
  \includegraphics[height=0.25\textwidth,width=0.3\textwidth]{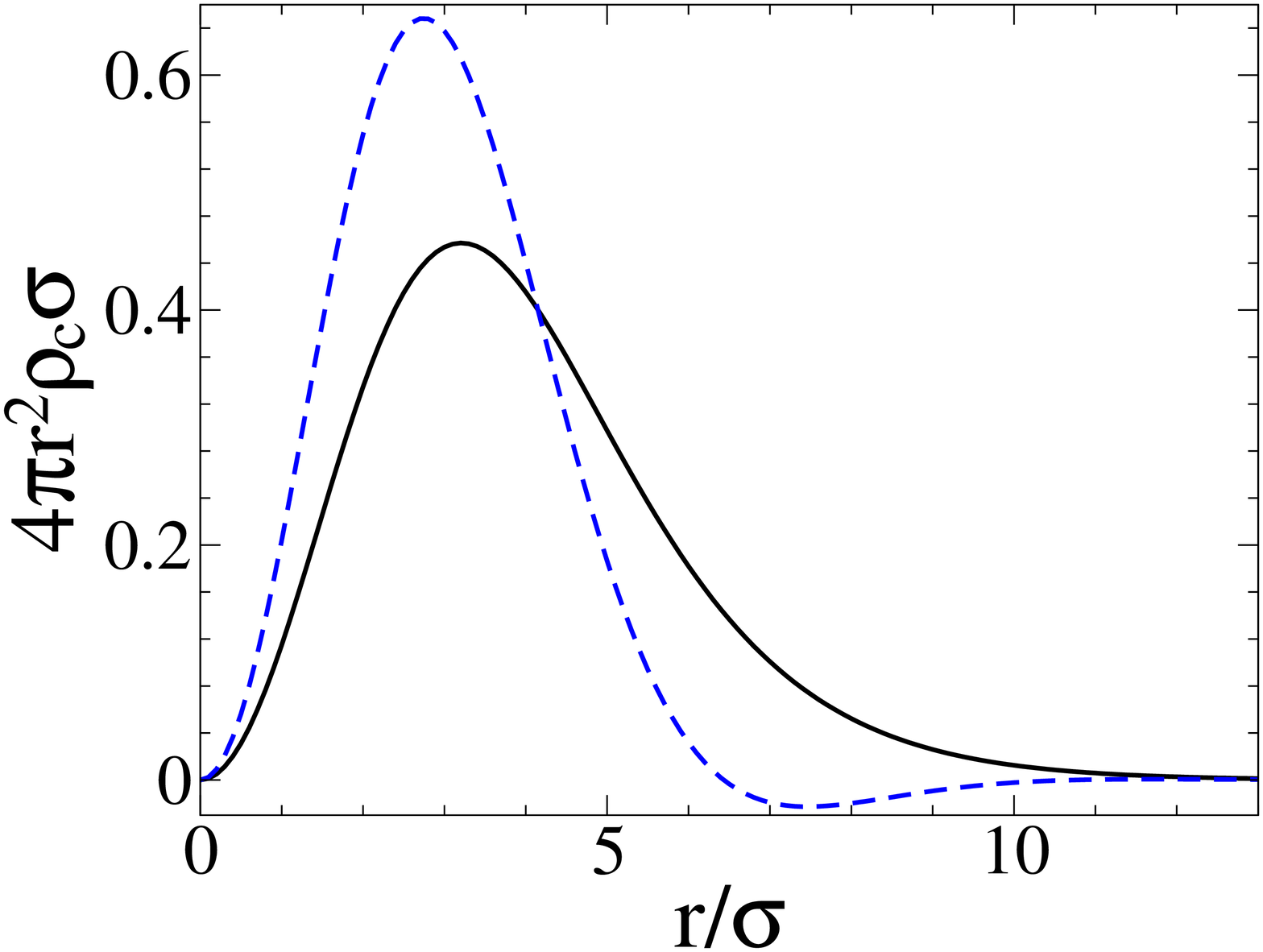}\\
\end{tabular}
 \end{center}
\caption{Charge density profiles, $4\pi r^2\rho_c(r)$, around a fixed particle  
for $\kappa_D\sigma=0.6$ (before the crossover, solid line) 
and $\kappa_D\sigma=1$ (after the crossover, dashed line).  The Kirkwood crossover is at $\kappa_D\sigma=e^{-1/2}\approx0.607$}
\label{fig:rho_c}
\end{figure}

In Fig. (\ref{fig:rho_c}) we plot charge density profiles generated by a fixed ion (but excluding a charge density
of that ion) and obtained from the Fourier transform of 
\be
\rho_c(k) = -2c_s w^2(k)\phi(k)
\ee
which in real space yields, using the residue theorem, 
\be
\rho_c(r) = -\frac{\kappa_D^2}{4\pi}\sum_{n=-\infty}^{\infty}\frac{e^{ik_n r}}{r} \frac{e^{-3k_n^2\sigma^2/2}}{1-k_n^2\sigma^2}.
\ee
\graphicspath{{figures/}}

\section{The Full mean-field}
\label{sec:sec4}
 
Having determined the screening parameters and the location of the Kirkwood crossover, 
we consider next the full nonlinear mean-field.  We are interested in the region before the Kirkwood 
crossover, $\kappa_D\sigma<e^{-1/2}$, where we determine the nonlinear renormalization of an effective 
surface charge.  We still consider Gaussian distributed ions.  

Before considering smeared-out ions, 
we first review some results for point-ions, 
whose mean-field description corresponds to the standard Poisson-Boltzmann equation,
which for the wall model is given by 
\be
\phi''(x) = \kappa_D^2\sinh\phi(x) - 4\pi\lambda_B\sigma_c\delta(x).  
\label{eq:PB1D}
\ee
After adapting the Debye screening length $\kappa_D^{-1}$ as a length scale (a dimensionless 
length is $y=\kappa_D x$) the above equation becomes  
\be
\phi''(y) = \sinh\phi(y) - \bigg(\frac{4\pi\lambda_B\sigma_c}{\kappa_D}\bigg)\delta(y).
\label{eq:SPB}
\ee
It now becomes clear that the functional form of $\phi(y)$ depends on a single parameter 
$4\pi\lambda_B\sigma_c\kappa_D^{-1}$.
Within the linear regime given by 
\be
\phi''_{\rm lin}(y) = \phi_{\rm lin}(y) - \bigg(\frac{4\pi\lambda_B\sigma_c}{\kappa_D}\bigg)\delta(y),
\label{eq:pb_lin}
\ee
the solution is 
\be
\phi_{\rm lin}(x) = \bigg(\frac{4\pi\lambda_B\sigma_c}{\kappa_D}\bigg)e^{-\kappa_D x}.  
\label{eq:phi_lin}
\ee

The nonlinear contributions do not renormalize the screening parameter, and the only parameter that is modified
is the magnitude of the far-field decay, or, as mentioned before, the "effective" surface charge $\sigma_c^{\rm eff}$, 
which is obtained by fitting the far-field potential to the functional form
\be
\phi(x)\approx \bigg(\frac{4\pi\lambda_B\sigma_c^{\rm eff}}{\kappa_D}\bigg) e^{-\kappa_D x},
\ee
or in a shorter form
\be
\phi(x)\approx A e^{-\kappa_D x}, 
\ee
where $A$ is a single fitting parameter.  Because Eq. (\ref{eq:SPB}) for point-ions admits an analytical solution, 
\be
\phi(y) = 2\log\bigg[\frac{4+A e^{-y}}{4-A e^{-y}}\bigg], 
\ee
which far away from a charged surface reduces to
\be
\phi(x) = A e^{-\kappa_D x} + O(e^{-3\kappa_D r}), 
\ee
the expression for $A$ is given by
\be
A = 4\Bigg(\sqrt{1+\bigg(\frac{2\kappa_D}{4\pi\lambda_B\sigma_c}\bigg)^2}-\frac{2\kappa_D}{4\pi\lambda_B\sigma_c}\Bigg),
\label{eq:A}
\ee
where the boundary conditions $\phi'(0) = -4\pi\lambda_B\sigma_c$ were used.

\begin{figure}[h] 
 \begin{center}%
 \begin{tabular}{rr}
  \includegraphics[height=0.25\textwidth,width=0.33\textwidth]{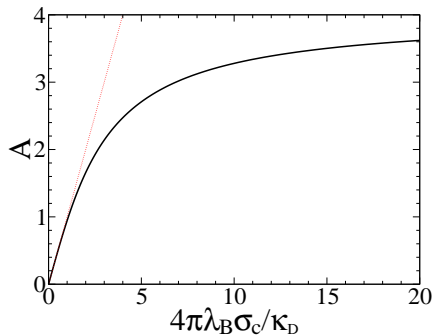}\\
\end{tabular}
 \end{center}
\caption{The coefficient $A=4\pi\lambda_B\sigma_c^{\rm eff}\kappa_D^{-1}$ as a function of $4\pi\lambda_B\sigma_c/\kappa_D$ 
obtained by fitting the far-field potential
to a functional form $Ae^{-\kappa_D x}$.  The dotted line corresponds to $A$ from a linear solution in 
Eq. (\ref{eq:phi_lin}), $A_{\rm lin}=4\pi\lambda_B\sigma_c/\kappa_D$.  }
\label{fig:sigma_point}
\end{figure}

In Fig. (\ref{fig:sigma_point}) we plot the coefficient $A$ 
as a function of $4\pi\lambda_B\sigma_c\kappa_D^{-1}$.  
The linear regime (indicated by a dotted line) breaks down already around $4\pi\lambda_B\sigma_c\kappa_D^{-1}\approx 1$, 
where the nonlinear contributions reduce the effective surface charge, eventually leading to saturation of $A$. 
Saturation implies that a charged surface no longer releases counterions into a solution 
but keeps them as part of a "dressed surface".  Apart from saturation, however, there is no charge inversion, which
for point-ions requires a more elaborate theory
\cite{Tellez05,Tellez06,Samaj06}.  
\graphicspath{{figures/}}

We consider next smeared-out ions.  The modified Poisson-Boltzmann equation in Eq. (\ref{eq:MPB}) for the wall geometry becomes
\ba
\phi''(y) \!\!&=&\!\! 
 (\kappa_D\sigma)^2\!\!\int \!\!d{\bf r}' w({{\bf r}},{{\bf r}}')\sinh\!\bigg[\!\!\int \!\!d{\bf r}'' w({\bf r}',{\bf r}'')\phi(y'')\bigg]\nonumber\\
&-& (4\pi\lambda_B\sigma_c\sigma)\delta(y) 
\label{eq:PB1Da}
\ea
where the unit of length is taken to be the size of an ion, $\sigma$ (where $y=x/\sigma$ and
${\bf r}\equiv {\bf r}/\sigma$ is a vector in reduced units).  
The functional form of $\phi(y)$ now depends on two parameters, $\kappa_D\sigma$ and $4\pi\lambda_B\sigma_c\sigma$, 
resulting in a more complicated solution than that for point-ions.  This additional degree of freedom provides an alternative 
route to charge inversion without contributions from correlations.

\begin{figure}[h] 
 \begin{center}%
 \begin{tabular}{rr}
  \includegraphics[height=0.25\textwidth,width=0.33\textwidth]{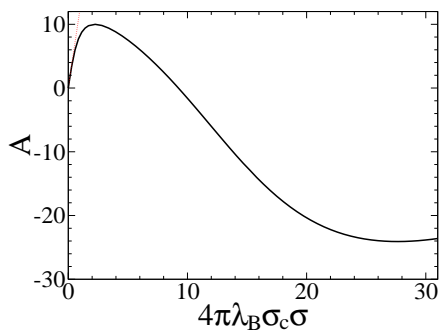}\\
\end{tabular}
 \end{center}
\caption{Coefficient $A$ (determined by matching the far-field behavior to the functional form $Ae^{-\kappa_0 x}$) as 
a function of a bare surface charge.  The plot is for $\kappa_D\sigma=0.33$, prior to the onset of oscillations.  
The system is a wall model and ions are Gaussian smeared-out ions.  The straight dashed line follows the linear result.  }
\label{fig:sigma}
\end{figure}

In Fig. (\ref{fig:sigma}) we plot the coefficient $A$ as a function of $4\pi\lambda\sigma_c\sigma$ prior
to the Kirkwood line, for Gaussian smeared-out ions.  $A$ was obtained by fitting a far-field potential to the functional 
form $Ae^{-\kappa_0 x}$, where
\be
\kappa_0\sigma = {\rm Re}\bigg[\sqrt{-W_0\big(-\kappa_D^2\sigma^2\big)}\bigg].  
\ee
As compared with a similar plot for point-ions in Fig. (\ref{fig:sigma_point}), the nonlinear renormalization 
of $A$ is considerably stronger.  After an incipient growth $A$ reaches a maximum at 
$4\pi\lambda\sigma_c\sigma\approx 2.3$ then at $4\pi\lambda_B\sigma_c\sigma\approx 9.1$ it changes sign, 
indicating the onset of charge inversion.  The inverted effective charge, however, does not increase indefinitely, 
and $A$ attains a minimum 
at $4\pi\lambda\sigma_c\sigma\approx 27.7$ after which it approaches zero for the second time, leading 
eventually to a subsequent charge inversion.  
\graphicspath{{figures/}}

\begin{figure}[h] 
 \begin{center}%
 \begin{tabular}{rr}
  \includegraphics[height=0.25\textwidth,width=0.33\textwidth]{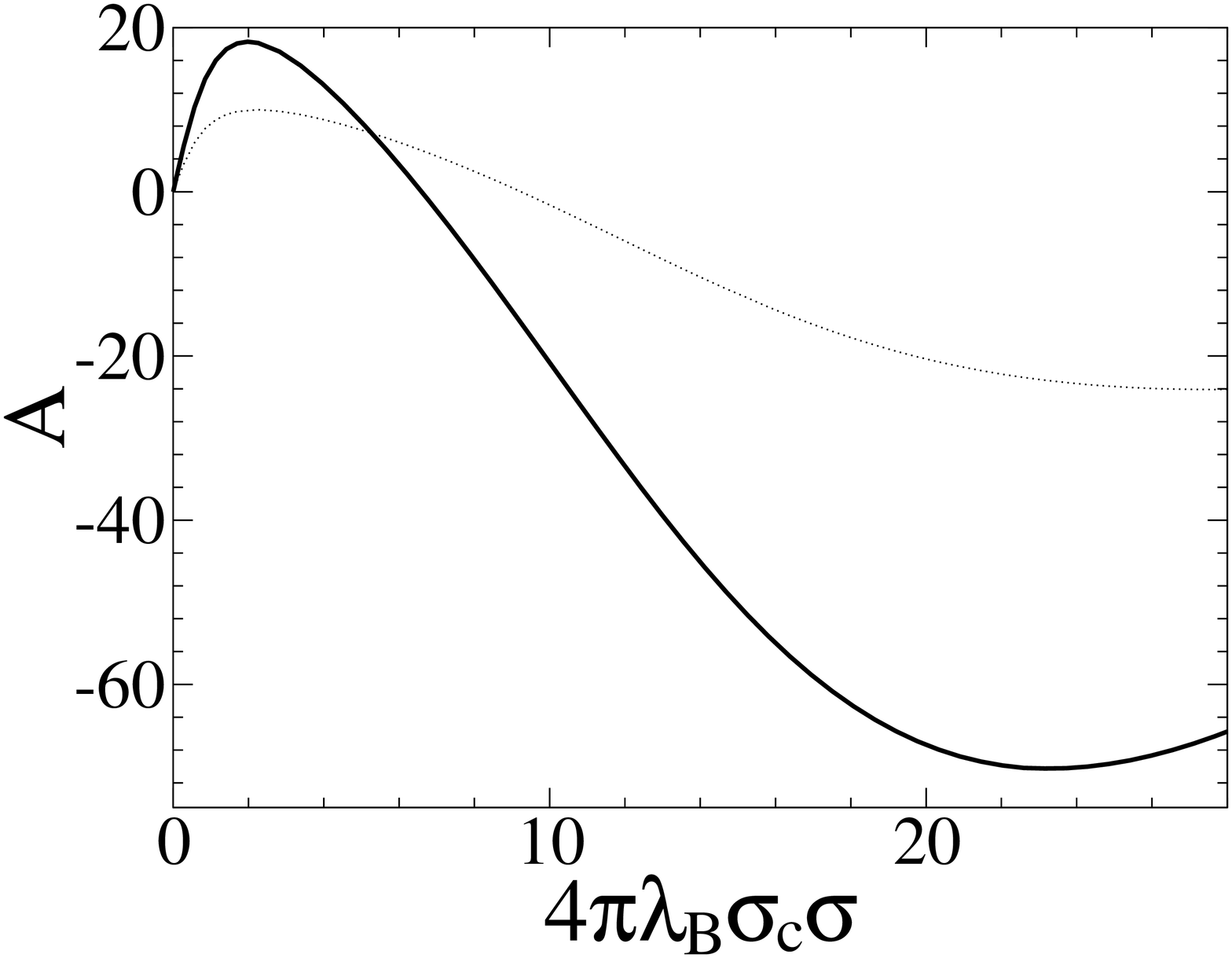}\\
\end{tabular}
 \end{center}
\caption{As in Fig. (\ref{fig:sigma}) but  
for $\kappa_D\sigma=0.47$ that is prior to the onset of oscillations.  
The dashed line is for comparison and corresponds to $\kappa_D\sigma=0.33$ in Fig. (\ref{fig:sigma}).  }
\label{fig:sigma2}
\end{figure}

\begin{figure}[h] 
 \begin{center}%
 \begin{tabular}{rr}
  \includegraphics[height=0.25\textwidth,width=0.33\textwidth]{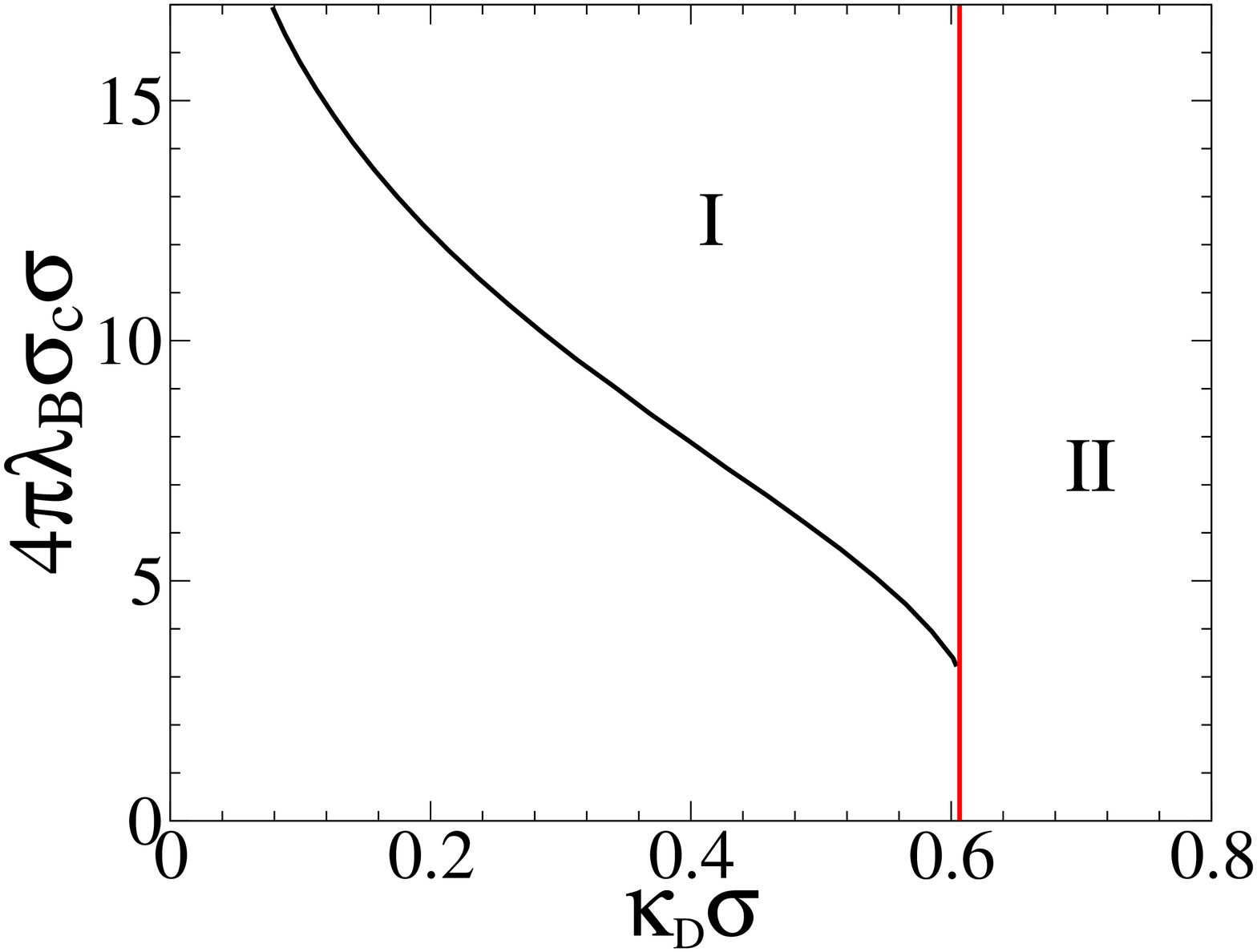}\\
\end{tabular}
 \end{center}
\caption{A diagram demarcating regions of charge inversion for Gaussian smeared-out ions.  The region $II$ corresponds 
to a region beyond the Kirkwood crossover, where charge inversion occurs by virtue of an oscillating charge density around zero.  
The region $I$ corresponds to charge inversion, before the Kirkwood
crossover, due to nonlinear renormalization of the coefficient $A$. }
\label{fig:top}
\end{figure}

In Fig. (\ref{fig:sigma2}) we show a similar plot but for $\kappa_D\sigma=0.47$.  
Apart from the increased magnitude there is also a shift in the position of charge inversion.  
\graphicspath{{figures/}}
In Fig. (\ref{fig:top}) we construct a diagram 
in the $(\kappa_D\sigma,4\pi\lambda_B\sigma_c\sigma)$ plane and demarcate the regions where charge 
inversion is possible.  
The region $II$ represents the region beyond the Kirkwood crossover, where charge inversion occurs by virtue of 
an oscillating profile.  Prior to the Kirkwood line is the region $I$, where charge inversion
is the outcome of nonlinear renormalization of an effective charge.  
\graphicspath{{figures/}}

\begin{figure}[h] 
 \begin{center}%
 \begin{tabular}{rr}
  \includegraphics[height=0.25\textwidth,width=0.35\textwidth]{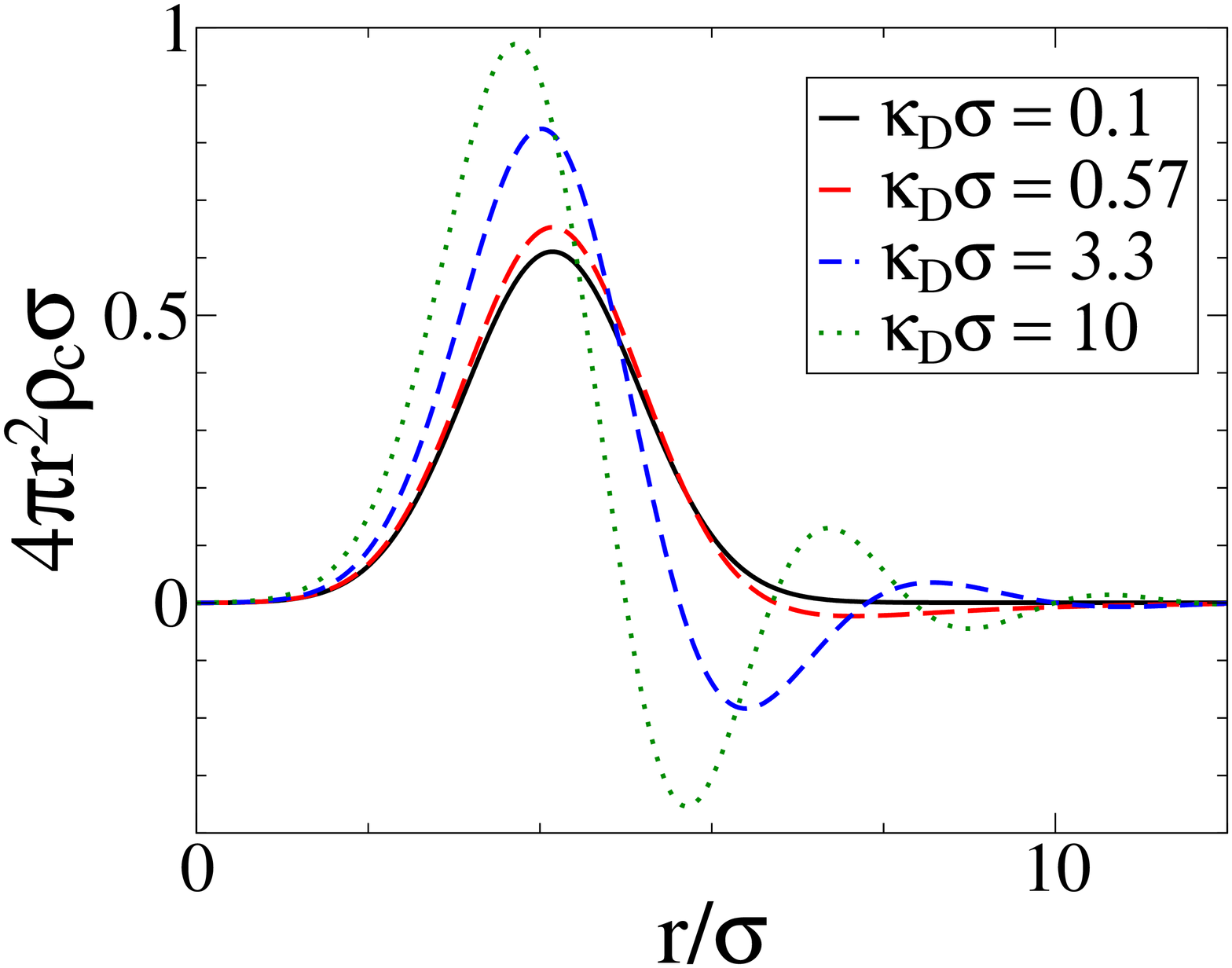}\\
\end{tabular}
 \end{center}
\caption{Charge density profile for a wall model obtained from a full mean-field theory for Gaussian smeared-out ions.  The results are for 
$4\pi\lambda_B\sigma_c\sigma=14$ and for three different values of $\kappa_D\sigma$:  $0.1$, $0.57$, $3.3$, and $10$.  
See Fig. (\ref{fig:top}) to locate these points in the $(\kappa_D\sigma,4\pi\lambda_B\sigma_c\sigma)$ plane.  }
\label{fig:rho_c2}
\end{figure}

As in the region I charge inversion occurs only through renormalization of an effective charge, and in a
low part of the region II (where nonlinear contributions are still weak) by virtue of an oscillating profile, the 
situation becomes more complex when the two contributions become significant.  
In Fig. (\ref{fig:rho_c2}) we plot charge density profiles for $4\pi\lambda_B\sigma_c\sigma\approx 14$
for different values of $\kappa_D\sigma$ corresponding to different points in the diagram in Fig. (\ref{fig:top}).
The curvature for $\kappa_D\sigma=0.57<e^{-1/2}$ exhibits charge inversion without oscillations.  For $\kappa_D\sigma=10$
oscillations become a dominant feature of the profile.  Here oscillations grow into
a full fledged layering of a charge density, a situation that is akin to the layering of oppositely charged polyelectrolytes 
as they become adsorbed onto a charged surface \cite{David03,Borkovec14}.  
\graphicspath{{figures/}}

\begin{figure}[h] 
 \begin{center}%
 \begin{tabular}{rr}
  \includegraphics[height=0.25\textwidth,width=0.3\textwidth]{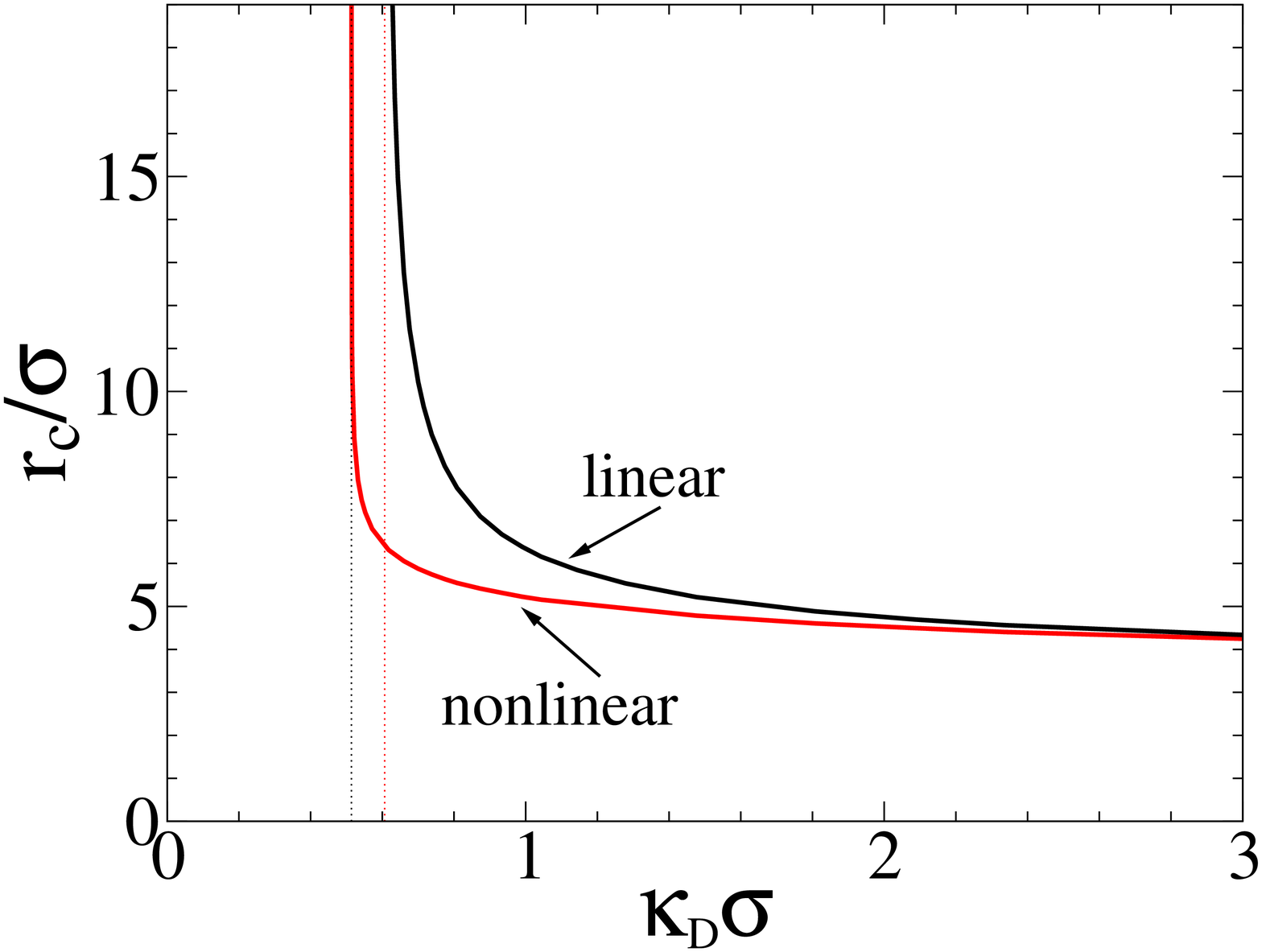}\\
\end{tabular}
 \end{center}
\caption{The shortest distance from a wall at which a potential becomes zero, $\phi(r_c)=0$, as a function of 
$\kappa_D\sigma$ for a charged wall model, for a linear and full mean-field theory.  The results are for 
$4\pi\lambda_B\sigma_c\sigma\approx 5.65$.  }
\label{fig:rc3}
\end{figure}

To further characterize 
charge inversion, in Fig. (\ref{fig:rc3}) we plot the shortest distance from a wall at which a potential changes sign, $\phi(r_c)=0$.  
At the onset of charge inversion, as $\kappa_D\sigma$ approaches the point of inversion from above, $r_c\to\infty$.  
Within the linear theory, divergence coincides with the Kirkwood crossover. 
On the other hand, the nonlinear contributions of the full mean-field theory push the divergence beyond the Kirkwood 
crossover.  
\graphicspath{{figures/}}

\section{Dumbbell ions}
\label{sec:dumbbell}

To place smeared-out ions in a larger perspective, we consider dumbbell ions that consist of two point charges
spatially separated and intended to represent some class of organic ions used as DNA condensing agents or short 
stiff polyelectrolytes \cite{Pincus08,Bohinc08,Bohinc11,Bohinc14,Frydel16}.  These ions have been shown to effect
bridging, and therefore attraction, between two same charged plates.  As this phenomena, too, has been captured 
by the mean-field, automatically we know that correlations are not involved and spatial extension alone bears responsibility.  
Based on these previous findings, we expect dumbbells to undergo charge inversion similar to that for spherically
smeared-out ions.  In this case, however, we have the complication of orientation:  dumbbells with parallel to a charged 
surface orientations within the mean-field are indistinguishable from point-ions.  Only configurations deviating from parallel 
orientation can effect charge inversion.    

Below we develop the mean-field framework for a dumbbell model.  
The normalized distribution of a single dumbbell is
\be
w({\bf r}-{\bf r}',{\bf n}) = \frac{1}{2}\bigg[\delta({\bf r}-{\bf r}') + \delta({\bf r}-{\bf r}' - \sigma{\bf n})\bigg]
\ee
and depends on orientation ${\bf n}$, where ${\bf n}$ is the unit vector.  
If the mean-field orientation dependent density is
\be
\rho_i({\bf r},{\bf n})\sim c_i e^{-\beta q_i(\psi({\bf r}) - \psi({\bf r}+\sigma{\bf n}))/2}, 
\ee
then a number density is obtained by averaging over an angular degree of freedom, 
\be
\rho_i({\bf r}) = c_i e^{-\beta q_i\psi({\bf r})/2}\int d{\bf n}\, \frac{e^{-\beta q_i\psi({\bf r}+\sigma{\bf n})/2}}{4\pi}.
\ee
By considering a wall model the above expression becomes
\be
\rho_i(x) = c_i e^{-\beta q_i\psi(x)/2}\int_{-\sigma}^{\sigma} ds\, \frac{e^{-\beta q_i\psi(x+s)/2}}{2\sigma}.
\ee
Finally, for a symmetric electrolyte with charges $q_{\pm}=\pm e$ and a bulk concentration $c_s$, the
charge density becomes
\be
\rho_c(x) = -\frac{c_s e}{\sigma} \int_{-\sigma}^{\sigma} ds\, \sinh\bigg[\frac{\beta e\psi(x) + \beta e\psi(x+s)}{2}\bigg], 
\ee
and the modified Poisson-Boltzmann equation in its dimensionless version is
\be
\phi''(x) = \frac{\kappa_D^2}{2\sigma}\int_{-\sigma}^{\sigma}ds\,\sinh\bigg[\frac{\phi(x)+\phi(x+s)}{2}\bigg].  
\ee



\begin{figure}[h] 
 \begin{center}%
 \begin{tabular}{rr}
  \includegraphics[height=0.2\textwidth,width=0.25\textwidth]{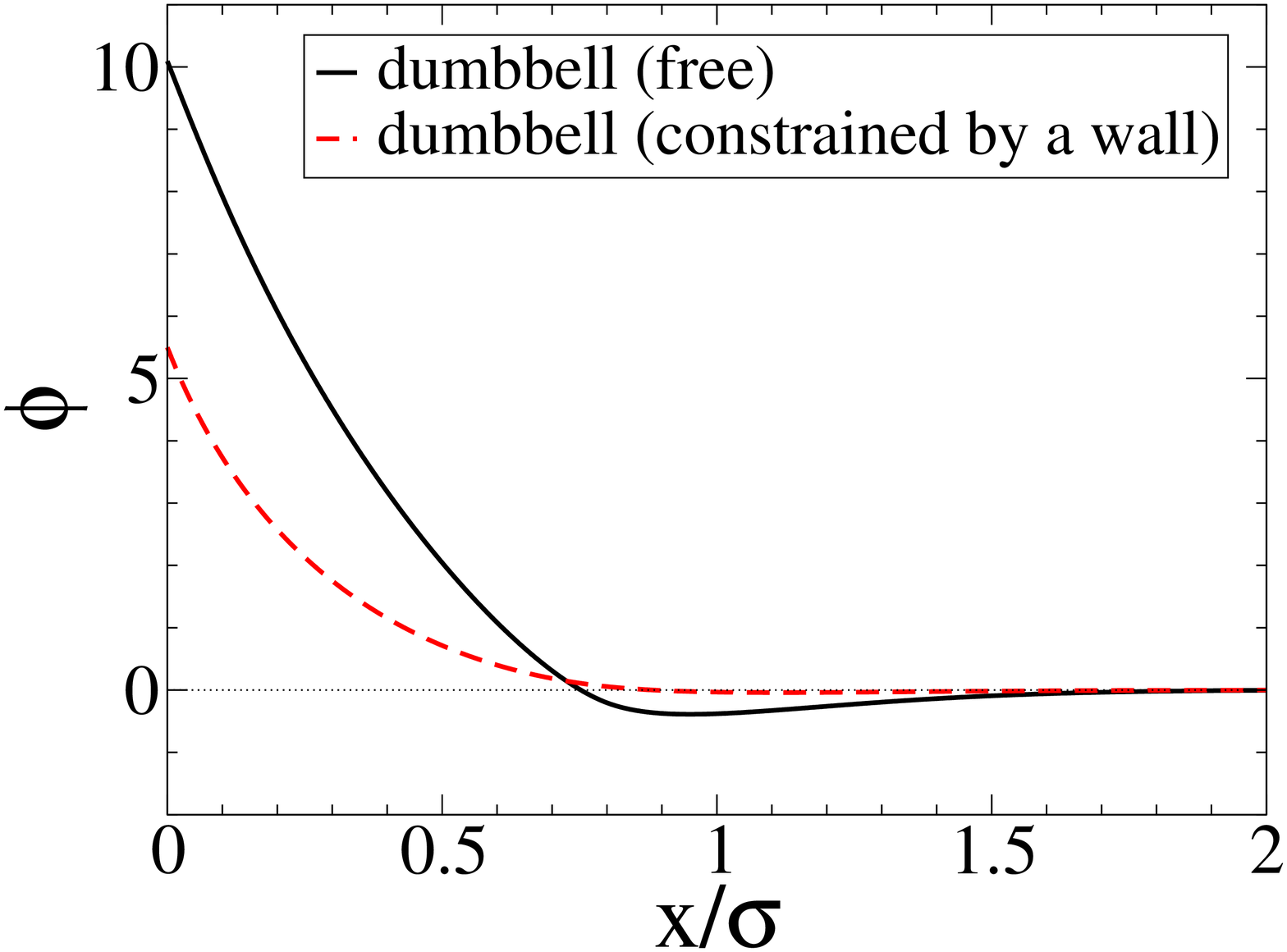}\\
  \includegraphics[height=0.2\textwidth,width=0.25\textwidth]{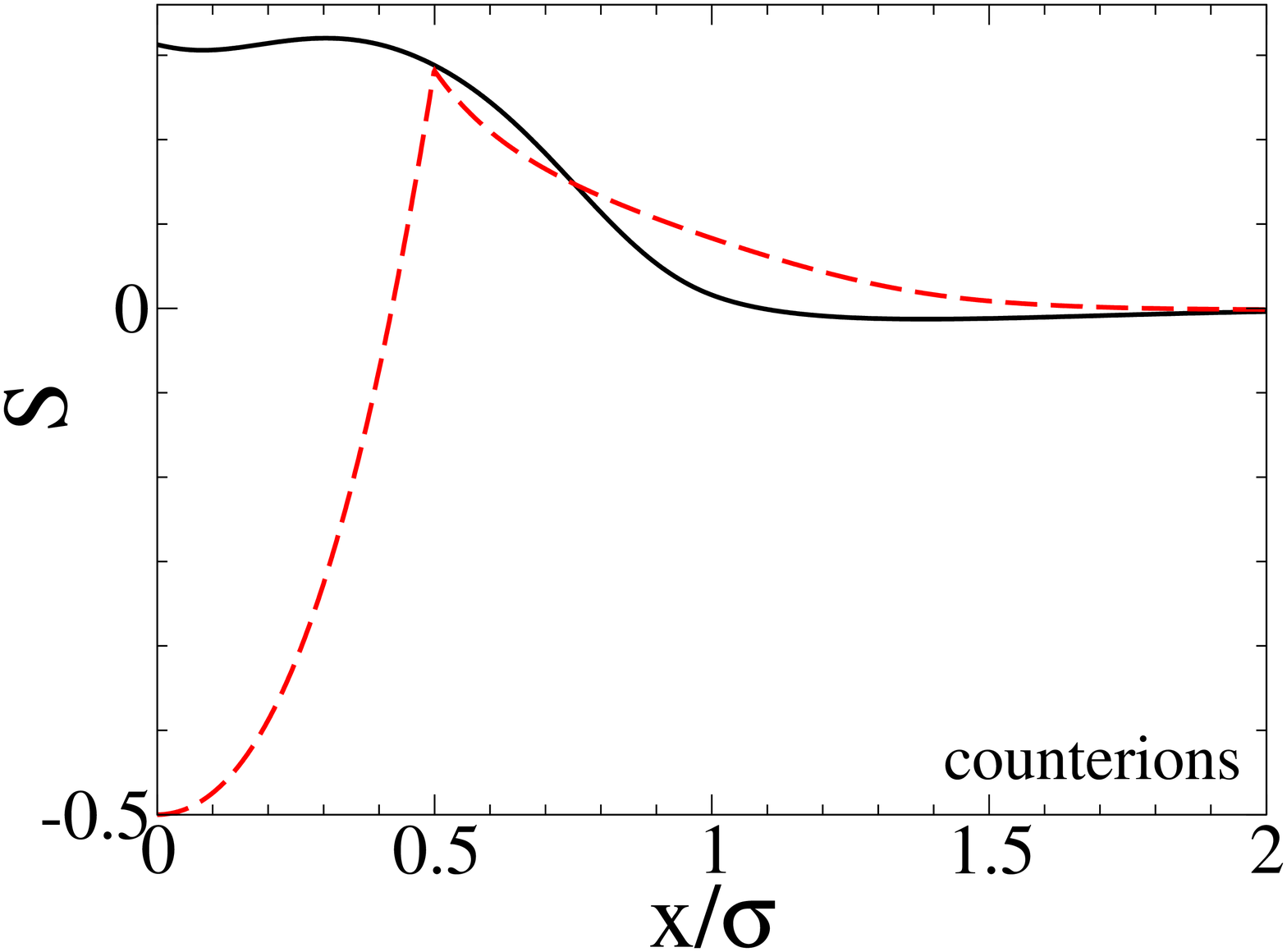}\\
\end{tabular}
 \end{center}
\caption{Electrostatic potential and orientation parameter $S$ as a function of a distance from a wall.  Relevant
parameters of are:  $\sigma=0.8\,{\rm nm}$, $\lambda_B=0.72\,{\rm nm}$, $\sigma_c=0.4\,{\rm Cm^{-2}}$, $c_s=1\,{\rm M}$.}
\label{fig:dumbbell}
\end{figure}

In Fig. (\ref{fig:dumbbell}) (a) we plot a potential for two types of angular behavior.  The black line 
shows results for ions whose orientation is unimpeded by a hard wall.  We call these ions "free".  
These ions clearly overscreen a charged surface and the potential becomes negative.  If, however, we 
allow a wall to limit possible orientations of nearby ions, (dashed line) we still find overscreening, however, 
considerably weaker.  
We can understand the situation by looking at the results in Fig. (\ref{fig:dumbbell}) (b) which plots the quantity 
\be
S = \frac{3\langle\cos^2\theta\rangle-1}{2},
\ee
as a function of a distance from a wall.  For perfect alignment $S=1$, for random orientation $S=0$, and $S<0$ indicates the 
preference for parallel orientations.  It is clear that if orientations become limited by a nearby wall, ions are forced into parallel 
orientations, lowering by the same toke the finite size effects.  
\graphicspath{{figures/}}

\section{Conclusion}
\label{sec:conclusion}

In the present work we have shown that charge inversion, normally linked to restructuring 
of counterions near a charged surface due to interactions between
surface charges and counterions, can occur through an alternative mechanism that depends on bulk properties 
of an electrolyte and exhibits qualitatively different behavior characterized by charge oscillations.  
The situation becomes more complex when the two 
mechanisms overlap and one has to distinguish between different contributions.  In such a situation
charge oscillations are no longer a far-field feature but produce full fledged layering of a charge density.  
The findings and conclusions of the present study need not be limited to smeared out charges but may apply to any type of 
electrolyte with Kirkwood crossover.  A primitive model would be one possible example.  Here the onset of oscillations 
is linked to the problem of hard-sphere packing \cite{Bing16}.  
A look into available literature reveals a number of studies having reported such an oscillating behavior \cite{Wu04}, 
without an explicit link to the Kirkwood crossover.

\begin{acknowledgments}
This research was supported by the Chinese National Science Foundation, the grant number 11574198.
Some computations were done using machines of the Laboratoire de Physico-Chime Th\'eorique, ESPCI.
\end{acknowledgments}



\end{document}